\documentclass[oneside]{article}
\usepackage[T1]{fontenc}
\usepackage{authblk}
\usepackage{amsmath}
\usepackage{graphicx}
\usepackage[english]{babel}
\usepackage{lmodern}
\usepackage[T1]{fontenc}
\usepackage[latin9]{inputenc}

\usepackage{graphicx}
\usepackage[unicode=true,
 bookmarks=false,
breaklinks=false,pdfborder={0 0 1},backref=false,colorlinks=false]
 {hyperref}
\usepackage[a4paper, total={210mm,297mm},margin=2.0cm]{geometry}

\usepackage{datetime}
\newdate{date}{17}{1}{2020}

\title{Relativistic antifragility} 

\author[1,2,3]{Sauro Succi \thanks{Electronic address: \texttt{s.succi@iac.cnr.it}; Corresponding author}}

\affil[1]{Center for Life Nano Science @Sapienza, Istituto Italiano di Tecnologia - 295 Viale Regina Elena, I-00161 Rome, Italy}
\affil[2]{Istituto per le Applicazioni del Calcolo CNR, Via dei Taurini 19, 00185 Rome, Italy}
\affil[3]{Institute for Applied Computational Science, Harvard John A. Paulson School of Engineering and Applied Sciences - Cambridge, MA 02138, USA}

\date{\displaydate{date}}

\begin{document}

\maketitle
 
\begin{abstract}
It is shown that the barbell distribution of a gas of relativistic
molecules above its critical temperature, can be interpreted as an 
antifragile response to the relativistic constraint of subluminal propagation.
\end{abstract}

\vspace{1cm}

\section{Introduction}
\label{intro}
In the recent years, the notion of anti-fragility, as introduced 
by Nassim Taleb \cite{TALEB}, has gained a boost of popularity 
across most walks of science and society.
Although to a physicist the term "anti-fragility" sounds a bit like  
a clever renaming of an old wine in a new bottle, the value of 
extending and applying the notion beyond the context of the natural
sciences, finance first, up to a philosophy of life, should not be underestimated.

In this short note, we wish to point out that relativistic kinetic theory, namely
the statistical theory of particles which obey Einstein's (special) relativity, shows
distinct signatures of anti-fragility, most notably the emergence of extreme (bell-bar) 
statistics in-lieu of standard "Mediocristian" gaussian distribution, in response 
to increasing thermal load.
The reason for such transition from Mediocristan to Extremistan, to borrow 
from Taleb's terminology is adamant: material particles cannot move 
faster than light.    
\section{Maxwell-Boltzmann statistics}
\label{sec:1}
It is well known that a gas of non-relativistic molecules at equilibrium obeys the
Maxwell-Boltzmann (MB) distribution (one spatial dimension for simplicity) \cite{BOL}:
\begin{equation}
\label{MB}
f_{MB}(v) = \frac{n}{(2\pi)^{1/2} v_T} e^{-\frac{(v-u)^2}{2v_T^2}}
\end{equation}
In the above $n$ is the gas number density (number of molecules per unit volume), 
$u$ is the macroscopic gas velocity and $v_T = \sqrt{\frac{k_BT}{m}}$ is the thermal
speed, $T$ being the temperature and $m$ the mass of the particle. 
According to kinetic theory, the macroscopic gas velocity $u$ 
coincides with the mean molecular velocity, namely: 
$$
u = <v> = \frac{1}{n} \int_{-\infty}^{+\infty} f v dv
$$
Note that the integral in velocity space runs from minus infinity to plus infinity since
classical mechanics sets no restrictions on the magnitude of the particle velocity.
Likewise, $mv_T^2/2=k_BT/2$ is the kinetic energy contained in the molecular fluctuations
$$
k_BT= <m(v-u)^2> = \frac{1}{n} \int f mv^2 dv
$$
In other words, the kinetic energy of the gas splits into
the sum of a macroscopic (mechanical) and a microscopic 
(thermal) components
$$
E_K = \frac{mu^2}{2} + \frac{m v_T^2}{2}
$$
In classical kinetic theory both terms are potentially unlimited.

A few comments are in order.

{\it First}, the mean velocity $u$ is also the most probable one, in the sense that
the MB distribution attains its peak value precisely at $v=u$.
This is typical conformistic-behaviour, most molecules 
"go with the flow", they move at the same speed as the average.

{\it Second}, such conformistic behaviour is fairly intolerant of outliers, namely
particles which move much faster or slower than the average are exponentially suppressed.
For instance, molecules moving at three thermal speed faster than average (FTA)
are suppressed by more than 1:1000, and at five FTA their number goes down 
to about one in a million! Extreme behaviour is suppressed in Mediocristan.  
Hence, "socially" speaking, the MB distribution speaks for a comfortable and 
stable world in which most individuals behave like the average and those who
don't are exponentially suppressed.
For good or for worst, this is the most stable statistics a gas of classical
molecules can achieve, as sharply pinpointed by Boltzmann's H-theorem, showing that
any different distribution is bound to converge to $f_{MB}$ in order to maximize
its entropy, basically a micrioscipic underpinning of the 
second principle of Thermodynamics.

{\it Third}, the MB distribution encodes the non-relativistic principle of 
Galilean invariance, i.e. the statistics does not depend on the 
absolute molecular velocity $v$, but on its speed relative 
to the average, $v-u$, also known as peculiar velocity, the 
one characterizing microscopic fluctuations. 
\section{Conservation constraints}
\label{sec:2}

The MB distribution maximizes entropy under 
the constraint of mass, momentum and energy conservation, i.e
\begin{eqnarray}
\int_{-\infty}^{+\infty} f_{MB} dv = n\\
\int_{-\infty}^{+\infty} f_{MB} v dv = nu\\
\int_{-\infty}^{+\infty} f_{MB} v^2 dv = nu^2 + n v_T^2
\end{eqnarray}
 
In Taleb's parlance, constraints are the "stressors", sources 
of stress, since they constrain the freedom of the system.

A few numbers won't hurt.
For air in standard conditions, 300 Kelvin degrees, the
thermal speed is close to $300$ meters per second (basically the sound speed).
If the gas is macroscopically at rest, $u=0$, the probability of finding a molecule
moving faster than one thermal speed, in either direction, is a sizeable 16 percent.
At two thermal speeds, the number goes down to about 2 percent, and at
above three thermal speeds, we are left with just one in thousands.
That means that in a sample of thousands molecules, on average, only 
one moves faster than $3 \times 300=900$ m/s.
By iterating the game, numbers get rapidly ridiculously small, at 
five thermal speeds, ($1500$ m/s) only one in a billion is left.
This speaks clearly for outlier suppression.

Now suppose that we set the gas in motion, with a substantial 
macroscopic speed, say $u=30$ m/s (100 Km/h, a pretty strong wind).  
How does the "molecular society" adjust to such stressor? 
The answer is fairly straightforward: by simply shifting 
the MB distribution towards positive values so that the 
new peak locates precisely at $v=u$. 
This breaks the left/right symmetry, the probability of finding 
a particle moving right at a given speed $+v$ is higher than the probability
of moving left with the same but opposite speed $-v$.
The probability of finding a molecule moving at $300$ m/s is $20/100$, slightly
larger than $16/100$ and the probability of moving at $-300$ m/s is $12/100$,
slightly smaller than $16/100$, but the distribution keeps 
the same bell-shaped form, symmetric around $u=30$ m/s.
The gas of molecules adjusts to the constraint without compromising any of the three
basic features described above, Mediocristan still rules.

Now suppose you keep the gas at rest, but increase its temperature instead,
say $600$ Kelvin instead of $300$.
How does the gas adjust to this thermal stressor?

Again the policy is adamant; by simply "broadening" its 
distribution, so that the average kinetic energy in the fluctuations 
is doubled, while the mean velocity is left unchanged. 
This means that now there are sixteen out of hundred molecules 
which travel at $600$ m/s instead of $300$ m/s. 
This is how the system manages to increase
its random energy content, but again, none of three distinctive 
features above are broken.
In particular, while the high velocity region
gets more populated, the most probable molecules remain those that
move at mean speed, in this case zero. 
The outliers are less suppressed, but they don't take over the conformists.
 
At this point, it is worth noting that the conformists contribute a little precious
nothing to the temperature constraint, since they carry no energy at all!
But they do contribute to the constraint that the gas should not move on average,
so they still have a role in this business.
This argument is flawed, but in classical mechanics the flaw remains 
silent, as we shall see shortly.

In classical kinetic theory you can play the game ad libitum, heat to the point
of making the sound speed equal to the speed of light, and the most probable molecules
will still be the ones at rest, although by an hardly appreciable extent, since the
MB distribution becomes utterly flat. 
The outliers are no longer such, granted, but the "conformist" are still there.

\begin{figure}
\centering
\resizebox{0.75\textwidth}{!}{%
  \includegraphics{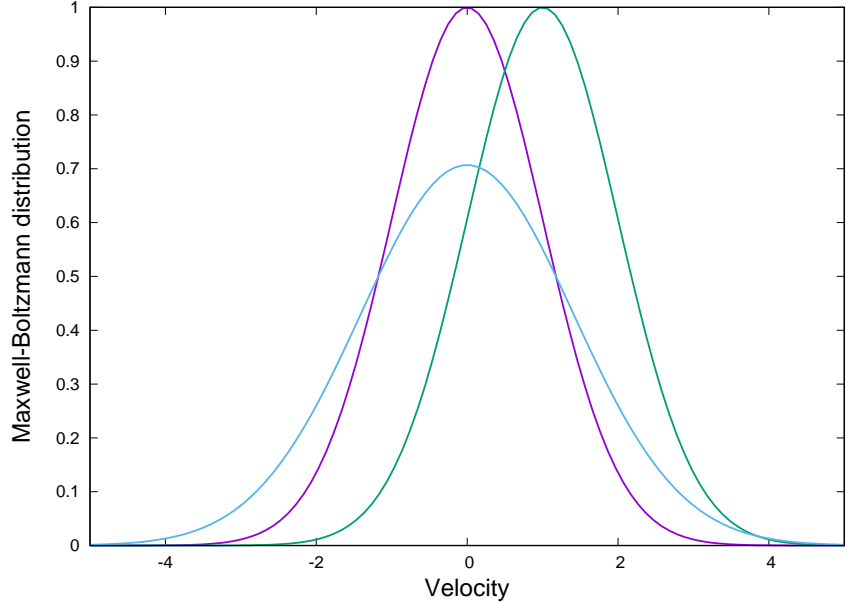}
}
\caption{The response of a non-relativistic gas at rest (center),with a net flow $u=0.1$ (right-shifted) and a temperature increase from $T=1$ to $T=2$ (center, broadened). By shifting and broadening, the MB distribution seamlessly adjusts to both flow and thermal constraints, respectively.}
\label{fig:1}       
\end{figure}
Since the outliers can afford unlimited speed, the system can absorb 
virtually any amount of thermal energy, by simply increasing their  
population at expense of the conformists. 
Yet, the thermal constraint can {\it always} be matched without destroying 
the conformist in the process, even though they contribute nothing to the energy budget!

The availability of unlimited speed provides the tolerance towards 
the zero-speed population of molecules.
\section{Relativistic kinetic theory}
\label{sec:3}
Next we move to relativistic molecules.

The relativistic analogue of the MB distribution, known as 
Maxwell-Juettner (MJ) distribution,reads as follows 
\cite{JUT,CERCI}:
\begin{equation}
f_{MJ}(v) = n A(z) \gamma_v^{3} e^{-z \gamma_v \gamma_u (1-uv)}
\end{equation}
where $\gamma_v = (1-v^2/c^2)^{-1/2}$ and
$\gamma_u = (1-u^2/c^2)^{-1/2}$ are the molecular and gas 
Lorentz factors, $z=mc^2/kT$ is the rest energy in thermal 
units and $A(z)$ a normalization prefactor \cite{CERCI}.
The MJ distribution follows directly from the expression of the relativistic
energy, $E^2=m^2 c^4 + p^2 c^2$, which in turn encodes 
Lorentz rather than Galilean in invariance. 
This means that the MJ depends no longer on the peculiar speed 
$v-u$, but on the (scalar in multi-dimensions) product $vu$. 

In addition, it obeys the boundary condition $f(v \to c) \to 0$, in 
compliance with the relativistic constraint $v \le c$. 

Crucial to the MJ statistics is the term $\gamma_v^{3}$, which stems 
from the metric transformation from the momentum distribution 
$f(p)$ to the velocity  distribution $f(v)$.
Such factor exhibits a cubic divergence in the limit $v/c \to 1$, thereby 
promoting depletion of the conformist population in favour of the outliers.
This divergence, in turn, results from the fact that while relativistic
velocities are bound, the corresponding momenta are not, which 

This divergence is tamed by the exponential term $e^{-\gamma_v}$
which enforces the boundary condition $f(v=c)=0$.
The MJ distribution reflects the basic competition between these 
two terms, whose outcome is strongly sensitive to temperature via 
the parameter $z$.

In the non-relativistic limits $v/c \ll 1$ and $1/z \to 0$, the 
MJ reduces to the MB distribution, hence it reacts 
to flow and thermal constraints in a similar way. 

In the genuinely relativistic limit $\gamma_v, \gamma_u \gg 1$,
however, the response is completely different.
To appreciate the point, let's consider again the case at rest, $u=0$.
It can be readily checked that upon increasing the temperature, i.e. $z \to 0$,
the MJ undergoes a depletion of the conformist region in favor of the emergence
of a barbell distribution, with two distinct peaks away from the origin.
More precisely, the transition from a MB-like unimodal to the bimodal 
barbell distribution occurs sligthly above a critical temperature 
$kT_{crit} \sim  mc^2$, namely below $z \sim 1$.
This stands in sharp contrast with the reaction of the MB distribution, 
which broadens indefinitely in response to the constraint of
increasing temperature, without ever developing any double-humped structure.
%
\begin{figure}
\centering
\resizebox{0.75\textwidth}{!}{%
  \includegraphics{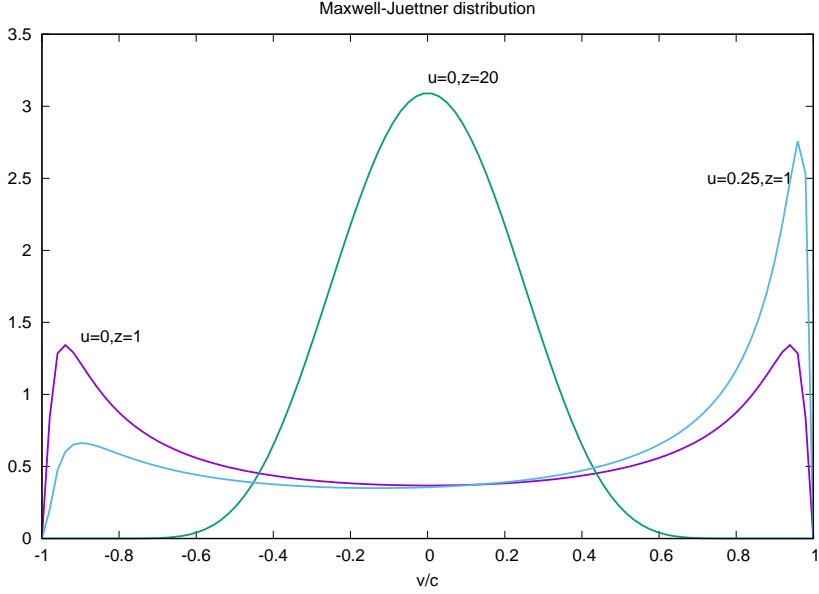}
}
\caption{The response of a relativistic gas at rest and $z=1$ and $z=20$, 
and to a net flow $u/c=0.25$ still at $z=1$.}
\label{fig:2}       
\end{figure}
Why such a different strategy?
Simply because the relativistic molecules cannot afford 
infinite-speed, the fast runners must nevertheless stop at $v=\pm c$. 
Hence, the only chance for the system to meet an increasing 
thermal constraint is to enhance the fastest-running populations as much as possible.
However, since the outliers can no longer afford infinite speed, a point comes
where the conformists are no longer sustainable, and above 
the critical temperature they begin to be suppressed.

But, didn't we say that the conformists are needed to comply with the 
net flow constraint, i.e. zero net motion for the case in point?

That is true, but now the system "realizes" that the same constraint 
can be fulfilled without the conformists, by simply keeping an exact balance 
between the left and right movers.
In other words, the relativistic barrier $v \le c$ exposes the 
"uselessness" of the comformists and the system
gets rid of them: that's Extremistan in full action!
  
But how about the case of a net macroscopic motion, say $u>0$?

Detailed inspection of the MJ distribution \cite{NUNO}, shows 
that this is accomodated by simply developing a positive bias on 
the right-moving hump and a negative one on the left-moving one.

This mechanism, known as "skweness", reflects Lorentz 
invariance, as opposed to the rigid shift of the non-relativistic 
case, which reflects Galilean invariance instead.

Quantitative analysis permits to compute the exact location 
of the humps and their width as a function of the macroscopic 
parameters $u/c$ and $z$, but does not add anything substantial
to the essence of the story.
And the essence is that, above a critical temperature, the
relativistic constraint $v \le c$ makes the conformist simply unsustainable.

The emergence of the barbell distribution then appears as quintessential 
antifragility, i.e an innovative survival strategy which was left
silent in the non-relativistic case, just due
to the availability of unlimited-speed runners.
\section{"Social" relativity}
\label{sec:4}
The "biological" interpretation of the MJ distribution is adamant: a survival strategy
against thermal constraints in a finite-resource (velocity) environment.
The social one, as usual, is a bit less straightforward.
The collapse of the conformist in relativistic gas is strongly conducive 
to the economic nosedive of the middle class in the modern 
"the winner takes it all" global economy.

With a big pinch of imagination, one might even posit that this 
is due to the mind-boggling acceleration of financial 
transactions, a sort of analogue of the relativistic limit $v \to c$.
In a world where data and algorithms can evaporate a lifetime savings 
at keystroke speed, quick-witted "influencers" and fake news, win hands 
down over "slow-paced" engineers, spelling doom for the real 
economy against the virtual one.

The second consideration is for the conformists, which relativity 
reveals in their true colors, basically as a disposable population. 
In human society, some individuals don't move because they
don't want to, but some others don't because they really can't. 
The antifragile policy is fair for the former but not for the latter.  

\section*{Acknowledgments}
This research leading has received funding from the European 
Research Council under the European Union's Horizon 2020 Framework 
Programme (No. FP/2014-2020)/ERC Grant Agreement No. 739964 (COPMAT).


\begin{thebibliography}{99}

\bibitem{TALEB} N. Taleb, 
Antifragile: Things That Gain From Disorder, 
Random House, NY (2012)

\bibitem{BOL} L. Boltzmann,
Lectures on Gas Theory, 
University of California Press, Berkeley, (1964)

\bibitem{JUT} F. Juettner,  
Annalen der Physik. 339 (5): 856-882 (1911), 

\bibitem{CERCI} C. Cercignani and G.M. Kremer, 
The Relativistic Boltzmann Equation: Theory and
Applications, Birkhauser, Berlin (2002).

\bibitem{NUNO} M. Mendoza, N. Arajuno, H. Herrmann and S. Succi, 
Sci. Rep, 2, 611, (2012)

\end{thebibliography}
\end{document}